\documentclass[12pt]{article}
\usepackage{graphicx}

\newcommand\pubdate{\today}

\def\Title#1{\begin{center} {\Large #1 } \end{center}}
\def\Author#1{\begin{center}{ \sc #1} \end{center}}
\def\Address#1{\begin{center}{ \it #1} \end{center}}

\newcommand\pubblock{\rightline{\begin{tabular}{l}  \\ 
         \pubdate  \end{tabular}}}
\newenvironment{Abstract}{\begin{quotation}  }{\end{quotation}}
\newenvironment{Presented}{\begin{quotation} \begin{center} 
             PRESENTED AT\end{center}\bigskip 
      \begin{center}\begin{large}}{\end{large}\end{center} \end{quotation}}
      
\newcommand{\beq}{\begin{eqnarray}}
\newcommand{\eeq}{\end{eqnarray}}
      
\begin{document}
\begin{titlepage}
 \pubblock
\vfill
\Title{Single inclusive particle production in pA collisions at forward rapidities: beyond the hybrid model}
\vfill
\Author{Tolga Altinoluk}
\Address{Theoretical Physics Division, National Centre for Nuclear Research, Pasteura 7, Warsaw 02-093,
Poland}
\Author{N\'estor Armesto\footnote{Speaker.}}
\Address{Instituto Galego de F\'{\i}sica de Altas Enerx\'{\i}as IGFAE, Universidade de Santiago de Compostela, 15782 Santiago de Compostela, Galicia-Spain}
\Author{Alexander Kovner}
\Address{Physics Department, University of Connecticut, 2152 Hillside Road, Storrs, CT 06269, USA}
\Author{Michael Lublinsky}
\Address{Department of Physics, Ben-Gurion University of the Negev, Beer-Sheva 84105, Israel}
\vfill
\begin{Abstract}
In this contribution we reconsider the calculation at next-to-leading order of forward inclusive single hadron production in $pA$ collisions within the hybrid approach. We conclude that the proper framework to compute this cross section beyond leading order is not  collinear factorization as assumed so far, but the TMD factorized framework.
\end{Abstract}
\vfill
\begin{Presented}
DIS2023: XXX International Workshop on Deep-Inelastic Scattering and
Related Subjects, \\
Michigan State University, USA, 27-31 March 2023 \\
     \includegraphics[width=9cm]{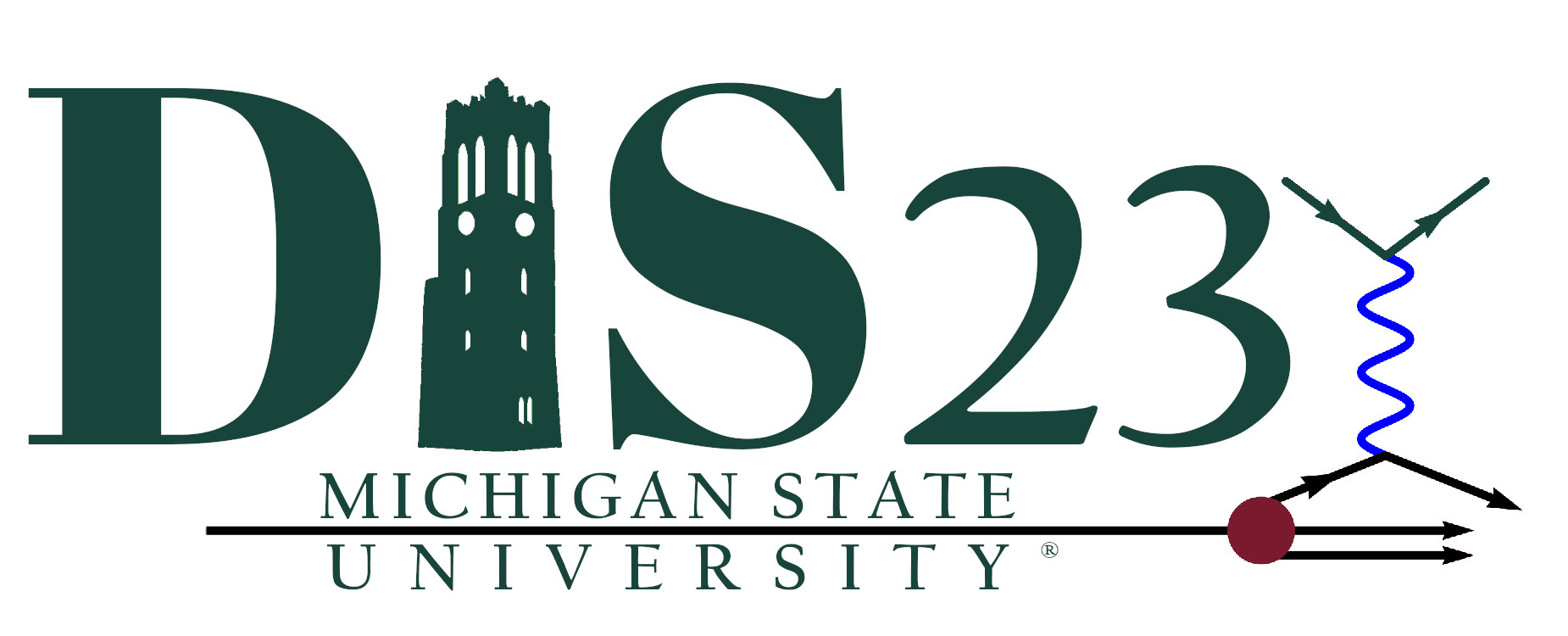}
\end{Presented}
\vfill
\end{titlepage}

\section{Introduction}

In the Color Glass Condensate (CGC), single inclusive particle production in the forward rapidity region in collisions between a dilute projectile, $p$, and a dense target, $A$, is usually addressed in the hybrid approach. The projectile is described through standard collinear parton densities. The projectile partons then scatter on the color field of the target modelled through averages of Wilson lines, to produce the final parton which subsequently hadronizes through standard collinear fragmentation functions.

The hybrid approach was formulated at the lowest order (LO) in~\cite{Dumitru:2005gt}. A partial next-to-leading order (NLO) calculation was performed in~\cite{Altinoluk:2011qy}, while the full NLO result appeared in~\cite{Chirilli:2011km,Chirilli:2012jd}. In this result, a collinear subtraction resulted in the DGLAP evolution of parton densities and fragmentation functions, while a rapidity cut-off to regularize the soft divergencies was required that led to the JIMWLK evolution equation, or its mean field version - the BK equation, for the averages of Wilson lines.

A numerical implementation of the NLO results in~\cite{Stasto:2013cha,Stasto:2014sea} showed that the cross section seemed to drop very fast with increasing transverse momentum and rather quickly became negative. Later, in~\cite{Altinoluk:2014eka} a restriction on the lifetime of the fluctuations of the projectile wavefunction (Ioffe time) was introduced that provided a soft cut-off and made evident the existence of additional NLO terms (resembling those in the BK evolution equation), which was also found in~\cite{Watanabe:2015tja} coming from kinematic restrictions. Such considerations alleviated the negativity problem but did not solve it completely. Further developments have been threshold~\cite{Liu:2020mpy,Xiao:2018zxf,Shi:2021hwx} and Sudakov~\cite{Shi:2021hwx} resummations. While these modifications resulted in  improved fits  to data, they seem rather ad hoc and there is no guarantee that they ensure positivity of the cross section at large transverse momentum.

In this contribution we revisit this NLO calculation following the formulation in~\cite{Altinoluk:2014eka}. We work in a frame in which the target is highly evolved, finding no need of additional BK evolution of the LO dipoles for physically motivated choices of factorization scales. We avoid any collinear or soft subtraction as previously done and conclude that the correct framework to resum the large logarithms that appear is not collinear but transverse momentum dependent (TMD) factorization for the projectile PDFs  and for the FFs. We will present here the results for the $q\to q\to H$ channel only. The complete results for all channels and full details can be found in~\cite{Altinoluk:2023hfz}.

\section{Transverse momentum distributions}

First of all we proceed to define our TMD PDFs\footnote{The definition of TMD FFs is completely analogous and, together with the generalization for gluons and several quark and antiquark species, can be found in~\cite{Altinoluk:2023hfz}.}. Several definitions of TMDs are used in the literature, differing predominantly in the way one treats the soft resolution scale, see, e.g.,~\cite{Collins:2011zzd,Boussarie:2023izj}, in addition to the process dependence of TMDs which requires, for different observables, the inclusion of different Wilson line factors in their definition, and to the nonperturbative aspects of TMDs. Here we use the simple intuitive perturbative definition, which is perfectly adequate for our purposes and resembles that used in the parton branching method~\cite{Hautmann:2017fcj,Martinez:2023azt}.

The unpolarized quark TMD PDF is defined perturbatively through its relation with the collinear PDFs as
\beq
\label{eq:tmdd1pp}
x{\cal T}_q(x,k^2;k^2; \xi_0)=\frac{g^2}{(2\pi)^3}\frac{N_c}{2}\int_{\xi_0}^{1}d\xi\frac{1+(1-\xi)^2}{\xi}\, \frac{x}{1-\xi}\,f^q_{k^2}\left(\frac{x}{1-\xi}\right)\frac{1}{k^2}\ .
\eeq
This  closely resembles the known perturbative relation between TMD and collinear PDFs in the large $k$ region~\cite{Collins:2011zzd,Boussarie:2023izj}. Here $\xi$ denotes the momentum fraction taken by the emitted gluon, the soft divergence in the gluon emission is regulated by the cutoff $\xi_0$, and the third argument in the TMD is the transverse resolution (factorization) scale.

The factorization scale dependence of the TMD PDFs is then given by the DGLAP-like equation
\beq
\label{eq:pdftmdmu}
x{\cal T}_q(x,k^2;\mu^2; \xi_0)&=&\theta(\mu^2-k^2)\, \Bigg[x{\cal T}_q(x,k^2;k^2; \xi_0) \\
&& \hskip 0cm -\frac{g^2}{(2\pi)^3}\frac{N_c}{2}\int_{k^2}^{\mu^2}\frac{\pi dl^2}{l^2}\,\int_{\xi_0}^{1}d\xi\frac{1+(1-\xi)^2} {\xi}x\,{\cal T}_q\left(x,k^2;l^2; \xi_0\right)\Bigg].\nonumber
\eeq
This equation can be interpreted in the following way: increasing the transverse resolution means that the number of quarks at a fixed transverse momentum decreases due to DGLAP splittings into quark-gluon pairs with higher longitudinal momentum given by the resolution scale.
Note that we regulate the soft divergence in gluon emissions by introducing the cut-off on momentum fraction, $\xi_0$\footnote{The definition of the longitudinal cutoff we use follows our earlier approach~\cite{Altinoluk:2014eka}, where we have limited the life time of the fluctuations by the Ioffe time cutoff. The resolution $\xi_0$ then depends on the virtuality $l$ of the gluon in the splitting as $\xi_0(l)=l^2/(xs_0)$, where $s_0$ is the Ioffe cutoff parameter.}. 

With these definitions, the collinear quark PDF, related to the quark TMD PDF via 
\beq
\label{eq:reltmdpdf}
xf^q_{\mu^2}(x)=\int_0^{\mu^2} \pi dk^2 \,x{\cal T}_q(x,k^2;\mu^2; \xi_0),
\eeq
satisfies the DGLAP evolution equations.

\section{The $q\to q\to H$ channel}
We start from the expressions derived in light-cone perturbation theory in~\cite{Altinoluk:2014eka} prior to any collinear or soft subtraction. Here we just outline the results and provide the final expressions; full details can be found in~\cite{Altinoluk:2023hfz}.

First, the real contribution can be reorganized into terms that yield the TMD PDFs and FFs as defined in (\ref{eq:tmdd1pp}), plus a genuine NLO remainder which shows no large soft or collinear logarithmic contribution.

Second, the virtual contribution can be manipulated to give the evolution of the TMD PDFs and FFs in the LO contribution as provided in (\ref{eq:pdftmdmu}) (expanded to order $g^2$), plus another NLO remainders which, for our choice of scales, see below, also show no large soft or collinear logarithmic contributions.

Working at large number of colors $N_c$, assuming translational invariance of the target averages of the ensembles of Wilson lines characterizing the target, and neglecting pieces that are either power suppressed or ${\cal O}(\alpha_s^2)$, our final expression for the quark channel, for production of a hadron with rapidity $\eta$ and transverse momentum $p$, reads
\beq
\label{eq:mytotal4}
&&\frac{d{\sigma}^{q\to q\to H}}{d^2p d\eta}=
S_\perp\int_{x_F}^1 \frac{d\zeta}{\zeta^2}\int d\xi \int d^2l \int d^2k\ 
{\cal F}_H^q\left(\zeta,l^2;\mu^2;\xi_0=\frac{\zeta \mu^2}{x_F s_0}\right)\\
&&\hskip 2cm \times\frac{x_F}{\zeta (1-\xi)} {\cal T}_q\left(\frac{x_F}{\zeta(1-\xi)},k^2;\mu^2; \xi_0=\frac{\zeta \mu^2}{x_F s_0}\right)\, {\cal P}(\xi,\zeta;k+l;p,s_0,\mu^2).\nonumber
\eeq
In this expression, $S_\perp$ is the transverse overlap area in the collision, ${\cal F}_H^q$ the TMD FF of a quark into a hadron $H$, and ${\cal T}_q$ the TMD PDF of a quark in the projectile.
The production probability reads
\beq
\label{eq:hardfactor4}
&&{\cal P}(\xi,\zeta;k+l;p,s_0,\mu^2)=\delta(\xi)\Bigg\{\int_qs\left(-(k+l)+\frac{p}{\zeta}\right)\left[1-\frac{(k+l)\cdot q}{q^2}\right]s\left(-q+\frac{p}{\zeta}\right)\nonumber \\
&& 
\hskip 0.8cm -2\frac{g^2}{(2\pi)^3} S_\perp \frac{N_c}{2}
\Bigg(\int_q \int_{\mu^2}^{\left(q-\frac{1}{\zeta}p\right)^2}\frac{d^2m}{m^2}\int_{m^2\zeta/(x_Fs_0)}^1 d\lambda \,\, \frac{1+(1-\lambda)^2}{\lambda} 
s\left(\frac{p}{\zeta}\right)\,s\left(q\right)\\
&&\hskip -0.8cm+\int_{m} \int_{0}^1 d\lambda \, \frac{1+(1-\lambda)^2}{\lambda}  \int_qs\left(\frac{p}{\zeta}\right)\,s(q)\,\left[\frac{\frac{p}{\zeta}(1-\lambda)-q-m}{(\frac{p}{\zeta}(1-\lambda)-q-m)^2}-\frac{\frac{p}{\zeta}-q-m}{(\frac{p}{\zeta}-q-m)^2}\right]\cdot
\frac{m}{m^2}\Bigg)\Bigg\}\nonumber\\
&&\hskip -0.8cm+\frac{g^2}{(2\pi)^3}\,\frac{N_c}{2}\int_m \frac{1+(1-\xi)^2}{\xi} \theta\left(\xi-\frac{m^2\zeta}{x_Fs_0}\right)\nonumber \\
&&\hskip -0.8cm \times \int_q s(m)s(q) \left[\frac{p/\zeta -m}{(p/\zeta -m)^2}-\frac{p/\zeta -(1-\xi)m}{(p/\zeta -(1-\xi)m)^2}\right]\cdot\left[\frac{p/\zeta -q}{(p/\zeta -q)^2}-\frac{p/\zeta -(1-\xi)q}{(p/\zeta -(1-\xi)q)^2}\right].\nonumber 
\eeq

As discussed in detail in~\cite{Altinoluk:2023hfz}, our expectation for the factorization scales for TMD PDFs and FFs are, respectively,
\begin{eqnarray}\label{thescales11}
&& \mu^2_T={\rm max}\left\{k^2,q^2,Q_s^2,\left(\frac{p}{\zeta}\right)^2\right\}\approx{\rm max}\left\{ (k+q)^2, Q_s^2,\left(\frac{p}{\zeta}\right)^2\right\}, \\
&& \mu^2_F=[(q+k)-p/\zeta]^2\approx {\rm max}\left\{(q+k)^2, (p/\zeta)^2\right\},\nonumber
\end{eqnarray}
with $Q_s$ the saturation scale of the target, $\zeta$ the  momentum fraction in the fragmentation function, $k$ the transverse momentum in the TMD PDF and $q$ the transverse momentum exchanged with the target.
On the other hand, the explicit calculations show that the main contribution to particle production arises from the kinematic regions $(k+q)^2\sim (p/\zeta)^2$ and $(k+q)^2\sim Q_s^2$.
Thus  both factorization scales become approximately $Q_s$ and $|p|/\zeta$ for $|p|\ll Q_s$ and $|p|\gg Q_s$, respectively. Therefore we expect the physically relevant choice to be
\beq\label{thescales1}
\mu_T^2=\mu^2_F=\mu^2={\rm max}\left\{Q_s^2,\left(p/\zeta\right)^2\right\}.
\eeq

\section{Conclusions}

We have revisited the calculation of single inclusive hadron production in $pA$ at forward rapidities at NLO within the hybrid approach. We have shown that beyond leading order the collinear resummation paradigm does not hold in this approach. In order to properly resum  large transverse logarithms at NLO one needs to work within the TMD factorization framework. 

The need to introduce TMDs is in fact quite  clear intuitively.
At high transverse momentum of the observed hadron, the naive parton model  picture of a low transverse momentum projectile parton that scatters with high momentum transfer off the target breaks down. We showed that this is one of the mechanisms for producing the high momentum hadron in the final state, but not the only one. The additional mechanism discussed in~\cite{Altinoluk:2011qy}, which amounts to the projectile parton acquiring large transverse momentum due to perturbative splittings in the projectile wave function, is at least equally important and should eventually guarantee the positivity of the cross section at large transverse momentum. It is clear that collinear factorization is physically inappropriate for the description of such a process.  We have shown in the present paper that the proper way to account for it, in the sense of proper resummation of  large transverse logarithms associated with perturbative splittings in the projectile wave function, is to view the parton as coming from the projectile TMD PDF with large transverse momentum. 

Also, we have identified for the first time the additional need of including the hard perturbative fragmentation of a low transverse momentum parton emerging from the scattering with the target. The naive collinear framework assumes that no hard momentum is produced in the fragmentation process. We showed here that this is not the case for the observable at hand, and 
that this process needs to be taken into account in order to resum the large logarithms.

\section*{Acknowledgements}

Special thanks are due to Guillaume Beuf who participated in early stages of this work.  
NA has received financial support from Xunta de Galicia (Centro singular de investigaci\'on de Galicia
accreditation 2019-2022), from the European Union ERDF, and from the Spanish Research State Agency under project PID2020-119632GB-I00.
AK is supported by the NSF Nuclear Theory grant 2208387. 
ML is supported by the Binational Science Foundation grant \#2021789.
This material is based upon work supported by the U.S. Department of
Energy, Office  of Science, Office of Nuclear Physics through the Saturated Glue (SURGE)
Topical Collaboration.
This work has been performed in the framework
of the European Research Council project ERC-2018-ADG-835105 YoctoLHC and the MSCA RISE 823947 "Heavy
ion collisions: collectivity and precision in saturation physics" (HIEIC), and has received funding from the European
Union's Horizon 2020 research and innovation program under grant agreement No. 824093.

\end{document}